\documentclass[letterpaper,twocolumn,pra,aps,showpacs,floatfix,superscriptaddress,longbibliography]{revtex4-1}
\usepackage[utf8]{inputenc}
\usepackage{bm}
\usepackage{multirow}
\usepackage{amssymb}
\usepackage{amsbsy}
\usepackage{amsmath,  amsthm, amsfonts, mathrsfs}
\usepackage{graphicx}
\usepackage{epsfig}
\usepackage{placeins}
\usepackage{physics}
\usepackage{trfsigns}
\usepackage{xfrac}
\usepackage[usenames,dvipsnames]{color}
\usepackage[colorlinks,linkcolor=blue,citecolor=blue,urlcolor=blue,breaklinks=true]{hyperref}

\usepackage{xr}
\makeatletter

\begin{document} 

 
\title{Stiffness of probability  distributions of work and Jarzynski relation for initial microcanonical and energy eigenstates}

 \author{Lars Knipschild}
 \email{lknipschild@uos.de}
 \affiliation{Department of Physics, University of Osnabr\"uck, D-49069 Osnabr\"uck, Germany}

 \author{Andreas Engel}
 \email{andreas.engel@uol.de}
 \affiliation{Institute of Physics, Carl von Ossietzky University of Oldenburg, D-26111 Oldenburg, Germany}

 \author{Jochen Gemmer}
 \email{jgemmer@uos.de}
 \affiliation{Department of Physics, University of Osnabr\"uck, D-49069 Osnabr\"uck, Germany}

\begin{abstract}
We consider closed quantum systems (into which baths may be integrated) that are driven, i.e., subject to time-dependent Hamiltonians.
 As a starting point we assume that, for  systems initialized in microcanonical states at some energies, the resulting probability 
 densities of work 
(work-PDFs) are largely independent 
of these specific initial energies. We show analytically  that this assumption of "stiffness", together with the assumption of an exponentially 
growing density of energy eigenstates, is sufficient but not necessary for the validity of the  Jarzynski relation (JR) for the 
above microcanonical initial states. This holds, even in the absence of microreversibility. To scrutinize the connection between stiffness and the JR for
 microcanonical initial states, we perform numerical analysis on systems comprising random matrices which 
may be tuned from stiff to nonstiff. In these examples we find the JR fulfilled in the presence of stiffness, and violated in its absence, which indicates a 
very close connection between stiffness and the JR. Remarkably, in the limit of large systems, 
we find the JR fulfilled, even for pure initial energy eigenstates. As this has no analogue  in classical systems, we consider it a genuine quantum  phenomenon.

\end{abstract}


\maketitle

\section{Introduction}

The long-standing question regarding whether, and in which way, closed finite quantum systems approach thermal equilibrium has recently 
gathered renewed attention. On the theoretical side thermalization and equilibration have been investigated e.g.\ for rather 
abstract settings \cite{Popescu2006,Goldstein2006,Reimann2008,Reimann2015,Eisert2015,Gogolin2016} and also for more 
specific condensed-matter type systems \cite{Rigol2008,Steinigeweg2013,Beugeling2014,Steinigeweg2014}. In these works major concepts
 are the eigenstate thermalization hypothesis (ETH) and typicality, both of which will also play certain roles in the  paper at hand. 
 The developments on experiments on ultra-cold atoms now allow for testing what have been merely theoretical results before; 
 see e.g.\  Ref.\  \cite{Reimann2016, balz17,tomkovic}.\\
Rather than just the existence of equilibration within closed quantum systems, lately  the very peculiarities of the  dynamical approach
 to equilibrium have moved to the center of interest \cite{Malabarba2014,Reimann2016}. 
Questions addressed in this context include limits on relaxation time scales and agreement of unitary quantum dynamics of closed quantum 
systems with standard statistical relaxation principles, such as Fokker-Planck equations \cite{thikonenkov13, ates12,Niemeyer2013,Niemeyer2014}, 
or more general, standard stochastic processes \cite{Gemmer2014,Schmidtke2016}. But also the emergence of universal non-equilibrium behavior 
involving work and driven systems is under  discussion at present \cite{Miller2017}.

To a large extent universal non-equilibrium behavior may be captured by fluctuation theorems, see e.g.\  Ref.~\cite{Seifert2008} and references
 therein. The Jarzynski relation (JR), a general statement on work that has to be invested to drive processes also and especially far from equilibrium,
  is a prime example of such a fluctuation theorem. Many derivations of the JR from various starting grounds have been presented. These include classical
   Hamiltonian dynamics, stochastic dynamics such as Langevin or master equations, as well as quantum mechanical
starting points \cite{Seifert2008,Roncaglia2014, Haenggi2015, Seifert2008,Esposito2009,Jarzynski2011,cuendet06}. However, all these derivations 
(except for Ref. \cite{Cleuren2006}) assume 
that the system, that is acted on with some kind of ``force'', is strictly in a Gibbsian equilibrium state before the process starts.
 (The notion of ``the system'' here routinely includes the bath.) Thus, this starting point differs significantly from the progresses in
 the field of thermalization: There, the general features of thermodynamic relaxation are found to emerge entirely from the system itself
  without any necessity of evoking external baths or specifying initial states in detail. Clearly, the preparation of a strictly
   Gibbsian initial state requires the coupling to a (super-)bath prior to starting the process.

This situation renders  the question whether or not the standard JR also holds  for systems starting in other than Gibbsian states 
(e.g. micro-canonical states) rather exigent. Note that, other than for Gibbsain initial states, the answer to this question is expected to depend on 
specific properties of the considered systems.

In this context a property which we call "stiffness of work-distributions" has been suggested as a key ingredient for the validity  of the JR for 
microcanonical initial states in Ref. \cite{Cleuren2006}. In this pioneering work the validity of the JR is proven for classical systems initialized 
in microcanonical initial states given the systems feature stiffness and microreversibility. Moreover, for a classical Lorentz gas stiffness and the 
validity of the JR for microcanonical initial states are 
numerically demonstrated. 
Furthermore  the JR was found to hold for micro-canonical initial states for some quantum spin-models exhibiting stiffness in Ref. \cite{stiffness} in a 
numerical study. The present work extends this line of research in various directions: We examine the validity of the JR not only for microcanonical 
initial states but also for initial pure energy eigenstates, the latter is conceptually beyond the scope of Ref. \cite{Cleuren2006}. It is also important 
to note that stiffness is a sufficient but not a necessary condition for the validity of the JR, thus the practical relevance of stiffness is challenged. 
The numerical modelling in the paper at hand allows to 
address this practical relevance 
by means of an investigation of the validity of the JR in the presence of stiffness, as well as in its absence. The latter is, to our best knowledge,
so far missing in the literature. Furthermore the results in the current paper do not rely  on microreversibility.

The paper at hand is organized as follows:
In Sec. \ref{sec:workdistri} we introduce our basic hypothesis of probability density function of work (work PDF's) being largely independent of 
the respective energy for micro-canonical initial states. We call this property  {\it stiffness}. The validity of the JR for micro-canonical initial states
 is shown to
 follow from this assumption (together with the routinely applied assumption of an exponentially growing density of energy eigenstates).
With an additional assumption on the system dynamics which we call {\it smoothness} we derive the validity of the JR even for energy eigenstates.
In Sec. \ref{numerical_setup} we introduce our modelling, which is partly based on random matrices. In Sec. \ref{numerical_results} we provide numerical 
results 
for micro-canonical initial states indicating a very strong correspondence between the validity of the JR and stiffness of the system dynamics.
In Sec. \ref{app_finite_size_scaling} we numerically show  that also the aforementioned smoothness-assumption is fulfilled for our modelling in the
limit of large systems. This completes the demonstration of the existence of a class of systems which exhibit both, stiffness and smoothness and thus
fulfill the JR even for energy eigenstates. We close with a discussion.


\section{Stiffness and Smoothness of Work Pdf's and Jarzynski Relation for initial microcanonical states and energy eigenstates}\label{sec:workdistri}

The analysis at hand focuses exclusively on closed systems. While it is physically appropriate to interpret the examples in Sec. \ref{numerical_setup} in terms of ``considered system'' and ``environment'' or ``bath'', we technically treat the system+environment compound regardless of the coupling strength as one closed system. Thus, since there is no external source or sink of heat, any energy change of the full system is to be counted as work $W$ (for an overview over different perspectives, see e.g.\  Ref.\  \cite{ueberblickJanet}.) The measurement of the inner energy is described by a two point projective measurement scheme. In this respect we choose the same starting point as employed in derivations of the JR as described, e.g.\ , in Ref.\  \cite{Campisi2011} and references therein. However, while in Ref. \cite{Campisi2011} the
assumption of a canonical, Gibbsian initial state is of vital importance, we base our consideration on much larger classes of initial states of the full system. The central role which the assumption of strictly Gibbsian state plays in the afore mentioned works is replaced by the assumption of ``stiffness'' of the work-PDF's (as introduced in in Eq. (\ref{stiffness})).


We consider a system described by a time-dependent Hamiltonian $H(t)$ during the time $t \in [0,T]$, which induces a non-equilibrium process.

The corresponding unitary time-propagation operator ${\bf U}$ is defined by:
\begin{equation}
 {\bf U} := \mathcal{T} \exp\left(-i\,\int^T_0 H(t') dt'\right) ~~~,
 \label{eq:U}
\end{equation}
where $\mathcal{T}$ is the time-ordering operator and we tacitly set $\hbar =1$.

Let $\ket{i}$ be the eigenstates of $H(0)$ and $\ket{f}$ the eigenstates of $H(T)$. Let further $\epsilon_i$ and $\epsilon_f$ be the corresponding eigenvalues, respectively.
Starting from the initial state $\ket{i}$, $p_{f \leftarrow i}$ denotes the probability to make a transition into $\ket{f}$:
\begin{equation}
p_{f \leftarrow i} = \mathrm{Tr}(\ket{f}\bra{f}U\ket{i}\bra{i}U^\dagger)
\end{equation}
The average over the work-PDFs $\langle h(W) \rangle_W$ starting from an initial state $\rho(0)$ can be calculated for an arbitrary function $h(W)$ of the work $W$:
\begin{equation} \label{calc_avg_eigen}
\langle h(W) \rangle_W = \sum_{i,f} \mathrm{Tr}(\rho(0) \ket{i}\bra{i}) p_{f \leftarrow i} h(\epsilon_f - \epsilon_i)
\end{equation}
$\mathrm{Tr}(\rho(0) \ket{i}\bra{i})$ is the probability to find the system after the first projective measurement in the initial state $\ket{i}$ and $p_{f \leftarrow i}$ is the probability to make a transition from $\ket{i}$ to the final state $\ket{f}$.
The work performed during this transition is $W=\epsilon_f - \epsilon_i$.

One can easily show that these transition-probabilities $p_{f \leftarrow i}$ are doubly stochastic:
\begin{equation} \label{sumprop_eigen}
 \sum_i p_{f \leftarrow i} = \sum_f p_{f \leftarrow i} = 1
\end{equation}
In general these transition-probabilities vary from eigenstate to eigenstate. We thus define the probability $p_{F \leftarrow i}$ to transition from an eigenstate $\ket{i}$ into an energy-interval $E_F$:
\begin{equation}
 p_{F \leftarrow i} = \sum_{f|\epsilon_f \in E_F} p_{f \leftarrow i}, \quad E_n = [n \delta, (n+1)\delta], \quad n=I,F
\end{equation}
Here, $\delta$ is to be chosen large compared to the level spacing of the full system, but small compared to the involved energy scales of $E, W$.
{Note that $I$ and $F$ are integers used to address the initial ($E_I$) and final energy-intervals ($E_F$), respectively. This construction serves as a coarse-graining of the energy scale.}

In a similar way, we define the average probability to make a transition from an initial state $\ket{i}$ from the energy-interval $E_I$ into an energy-interval $E_F$:

\begin{equation} 
 p_{F \leftarrow I} = \sum_{i|\epsilon_i \in E_I} \frac{p_{F \leftarrow i}}{\Omega_I}
\end{equation}
$\Omega_I$ and $\Omega_F$ denote the number of eigenstates of $H(0)$ in the interval $E_I$ and of $H(T)$ in the interval $E_F$, respectively.
\begin{equation}
\Omega_n = \mathrm{Tr}(\Pi_n), \quad
\Pi_I = \sum_{i|\epsilon_i \in E_I} \ket{i}\bra{i}, \quad 
\Pi_F = \sum_{f|\epsilon_f \in E_F} \ket{f}\bra{f}
\end{equation}
Hence, $p_{F \leftarrow I}$ is the average over all $p_{F \leftarrow i}$ with $\epsilon_i \in E_I$.

Note that these transition-probabilities depend on the width $\delta$ of the final energy-interval.
Closely related to these transition-probabilities is the so-called work probability density function (work-PDF), which describes the probability to perform the work $W=(F-I) \delta$ starting from an initial energy $E = I \delta$.
\begin{equation}
P_{E}(W) = \frac{1}{\delta} p_{F \leftarrow I}
\end{equation}
The transition-probabilities and the work-PDFs are essentially the same, up to a constant rescaling-factor.
But in large systems these work-PDFs typically become independent of the concrete choice of $\delta$ \cite{stiffness}.

Starting from Eq. (\ref{calc_avg_eigen}), the average over the work-PDFs $\langle h(W) \rangle_W$ for a function $h(W)$, which does not vary significantly on the scale of $\delta$, can be calculated from $p_{F \leftarrow I}$:
\begin{equation} \label{calc_avg}
 \langle h(W) \rangle_W = \sum_{I,F} \mathrm{Tr}(\rho(0) \Pi_I) p_{F \leftarrow I} h(\bar{E_F} - \bar{E_I})
\end{equation}
$\bar{E_n} = n \delta$ is an approximation of the energies in the initial ($n=I$) interval $E_I$ and of the final ($n=F$) interval $E_F$, respectively.

From Eq. \ref{sumprop_eigen} we derive the following properties of $p_{F \leftarrow i}$ and $p_{F \leftarrow I}$:
\begin{equation}
\sum_i p_{F \leftarrow i} = \Omega_F
\end{equation}

\begin{equation} \label{sumprop}
\sum_I \Omega_I p_{F \leftarrow I} = \Omega_F
\end{equation}

Up to now we only defined various quantities and derived general statements, but did not make any assumptions.
We now come to the derivation of the JR for micro-canonical initial states. To begin with, we define the latter as 
\begin{equation}
\rho_\text{mc}^I(0) = \Pi_I \Omega_I^{-1}
\end{equation}
In order to derive the JR for micro-canonical initial states to make two assumptions.
First, we assume that the probability to make a transition from a state from the energy-interval $E_I$ into the energy-interval $E_F$ only depends on the difference of $F$ and $I$:
\begin{equation} \label{stiffness}
 p_{F \leftarrow I} = p(F - I)
\end{equation}
We call this assumption {\it stiffness}. This assumption can be also expressed in terms of work-PDFs $P_E(W)$. 
If these work-PDFs are independent of the initial energy $E$, then Eq. (\ref{stiffness}) is fulfilled.

{
Our second assumption states that the densities of states (DOS) of the initial $D_\mathrm{ini}(E_I) := \delta^{-1} \Omega_I$ and final Hamiltonian $D_\mathrm{fin}(E_F) := \delta^{-1} \Omega_F$ grow exponentially:
\begin{align}
\begin{split} \label{exp_dos}
 D_\mathrm{ini}(\bar{E_I}) &= Z_\mathrm{ini} \exp{\beta \bar{E_I}}, \\
 D_\mathrm{fin}(\bar{E_F}) &= Z_\mathrm{fin} \exp{\beta \bar{E_F}}
\end{split}
\end{align}
	Up to now $\beta$, $Z_\mathrm{ini}$ and $Z_\mathrm{fin}$ are just some positive real numbers.
In the discussion below (\ref{JR0}) these numbers are interpreted in terms of standard statistical thermodynamics.
}

Of course Eq. (\ref{stiffness}) and Eq. (\ref{exp_dos}), are not expected to hold for all energies $E$. Here we only require that these relations hold at least for an energy interval which is large enough to comprise almost the entire work-PDF.

To arrive at the JR for micro-canonical initial states, we start by calculating the average of $\exp{-\beta W}$ over the work-PDFs according to Eq. (\ref{calc_avg}).
\begin{align}
\begin{split}
\langle & \exp{-\beta W}\rangle_W \\
	       & = \sum_{I', F} \mathrm{Tr}(\rho_\text{mc}^I(0) \Pi_{I'}) p_{F \leftarrow I'} \exp{-\beta (\bar{E_F} - \bar{E_{I'}}} \\
	       & = \sum_{F} p(F - I) \exp{-\beta (\bar{E_F} - \bar{E_{I}}}
\end{split}
\end{align}
In the last step we evaluated the sum over $I'$ by using $\mathrm{Tr}(\Omega_I^{-1} \Pi_I \Pi_{I'}) = \delta_{I, I'}$ and used the stiffness-assumption Eq. (\ref{stiffness}). By substituting $F$ by $F' + I - I'$, while $I'$ is the new summation index and $F'$ an arbitrary but fixed integer, we get:
\begin{align} \label{JR0}
\begin{split}
\langle & \exp{-\beta W} \rangle_W = \\
	& = \sum_{I'} p(F' - I') \exp{-\beta (\bar{E_{F'}} - \bar{E_{I'}}} \\
	& = \frac{1}{\Omega_{F'}} \frac{Z_\mathrm{fin}}{Z_\mathrm{ini}} \sum_{I'} p_{F' \leftarrow I'} \Omega_{I'} = \frac{Z_\mathrm{fin}}{Z_\mathrm{ini}}
\end{split}
\end{align}
In the second step we used that the DOS of the initial and the final Hamiltonian exponentially grow according to Eq. (\ref{exp_dos}).
In the last step we used Eq. (\ref{sumprop}).

Eq. (\ref{JR0}) formally is a JR for the work PDF's obtained by starting from microcanonical initial states, with the temperature replaced by a parameter 
describing the exponential growth of the DOS of the full system. As such Eq.\ (\ref{JR0}) already represents the main result of the present section.
 Note that Eq.\ (\ref{JR0}) holds for arbitrary processes and its r.h.s. only contains static, process-independent model parameters.

Formally the JR could be fulfilled for microcanonical initial states, even if Eq. (\ref{stiffness}) and Eq. (\ref{exp_dos}) do not hold. In this sense these assumptions  are
 stronger than the validity of the JR, or to rephrase, these assumptions represent sufficient but not necessary conditions. This peculiarity will be 
 investigated in detail below

In an analogous way we can derive Eq. (\ref{JR0}) for initial energy eigenstates $\rho(0) = \ket{i}\bra{i}$ if we additionally assume that
\begin{equation} \label{smoothness}
 p_{F \leftarrow i} \approx p_{F \leftarrow I}
\end{equation}
holds for all $i \in \mathbb{N}$ with $\epsilon_i \in E_I$. This additional assumption means that the transition probabilities from an eigenstate $\ket{i}$ to an energy interval $E_F$ are smooth functions of the initial and final energy. We therefore call it {\it "smoothness"}. 
The validity of this assumption is investigated in Sec. \ref{app_finite_size_scaling} in a finite size scaling.

In order to demonstrate even closer analogy of Eq. (\ref{JR0}) with the standard JR, it remains to be explained in which sense the r.h.s of Eq.\ (\ref{JR0}) may be  considered as the familiar r.h.s of the standard JR, $e^{-\beta \Delta F}$, where $F$ is the free energy. Such an identification would hold if 
\begin{equation}
 \label{eq:freeen}
 -\frac{\ln Z_\alpha}{\beta} \stackrel{?}{=} F_\alpha~~~.
\end{equation}
In order to judge whether or not Eq.\ (\ref{eq:freeen}) is justified, consider the logarithm of Eq.\ (\ref{exp_dos}),
\begin{equation}
 \label{eq:mustexpln}
	\ln  D_\alpha(U) =  \ln  Z_\alpha + \beta U~~~.
\end{equation}
The index $\alpha \in \{\mathrm{ini}, \mathrm{fin}\}$ signals, whether the equation refers to the initial or final Hamiltonian, respectively.
Moreover, the discrete average Energies $\bar{E_I}$ and $\bar{E_F}$ are replaced by the continous parameter $U$.

If one identifies, along the lines of Boltzmann's original approach, the entropy $S_\alpha$ as
\begin{equation}
 \label{eq:boltzent}
 \ln  D_\alpha := S_\alpha
 \end{equation}
(where we tacitly set $k_B =1$), one may convert Eq.\ (\ref{eq:mustexpln}) into
\begin{equation}
 \label{eq:freeident}
 -\frac{\ln Z_\alpha}{\beta} = U - \frac{S_\alpha}{\beta}.
\end{equation}

Note that, in accordance with Eq.~(\ref{exp_dos}), $\partial_U S_\alpha = \beta$, hence $\beta$ has the meaning of inverse temperature, and the r.h.s. of Eq.\ (\ref{eq:freeident}) is, accordingly the free energy $F$ as introduced in standard textbooks on phenomenological thermodynamics. In this sense  
Eq.\ (\ref{eq:freeen}) indeed holds, which entails the rewriting of Eq.\ (\ref{JR0}) in a form closer to the familiar one:
\begin{equation}
 \langle  e^{- \beta W} \rangle_E =e^{-\beta \Delta F}~~~,
 \label{jarzynski}
\end{equation}
where $\langle \cdots \rangle_E$ denotes the microcanonical expectation value corresponding to energy $E$. This concludes our consideration on the validity of a JR for microcanonical initial states under the assumption of stiff work-PDFs.

\section{Models and Driving Protocol} \label{numerical_setup}
With the following  numerical investigations we ascertain the pivotal relevance of stiff work-PDFs for the validity of the 
JR for microcanonical initial states.
We therefore introduce a model that is partly based on random matrices. 
Within this model we can control the stiffness of the resulting work-PDFs via a single parameter $\xi$. This allows us to observe the influence of 
stiffness on the JR for microcanonical initial states.

\begin{figure}
\includegraphics[width=0.4\textwidth]{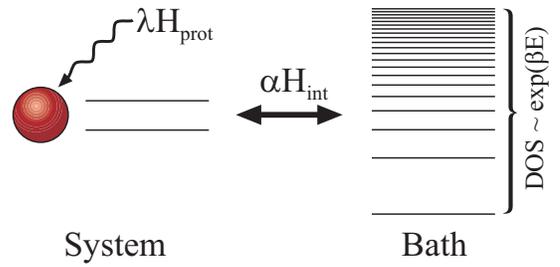}
\caption{
	Schematic structure of the numerical model. A two-level system is coupled via a random-interaction to a bath with an exponentially growing DOS.
	The structure of the interaction takes influence on the resulting work-PDFs.
}
       \label{fig_sys}
\end{figure}

We consider an isolated system comprising a relatively small subsystem (denoted as "sys", $H_\mathrm{sys}$) and a bigger part serving as heat bath 
(denoted by "bath", $H_\mathrm{bath}$).
Both parts may interact via $H_\mathrm{int}$. Finally, a time-dependent force periodically drives the system $H_\mathrm{prot}$. Concretely we 
choose the small subsystem to be a spin and the time dependent force to be a kind of microwave field such that the whole model allows for an 
interpretation in terms of a spin-resonance experiment with a finite lifetime of the spin excitation,  see Fig. \ref{fig_sys}. A very similar model 
(spin-GORM model) has previously been used to study relaxation in finite environments \cite{Esposito_Relaxation}

In detail the Hamiltonian of the full system reads:

\begin{equation}
H(t) = H_\mathrm{sys} + H_\mathrm{bath} + \alpha H_\mathrm{int} + \lambda H_\mathrm{prot}(t)
\end{equation}

The small subsystem is a simple two-level system, e.g. a spin-\sfrac{1}{2}-particle in a magnetic field $B_\mathrm{z}$.
The Hamiltonian of this subsystem is characterized as:
\begin{equation}
H_\mathrm{sys} \ket{E^\mathrm{sys}_j} = E^\mathrm{sys}_j \ket{E^\mathrm{sys}_j}, E^\mathrm{sys}_{1|2} = \mp \frac{B_\mathrm{z}}{2}
\end{equation}
$\ket{E^\mathrm{sys}_j}$ obviously denote the eigenstates of $H_\mathrm{sys}$. We chose $B_\mathrm{z} = 0.5$ throughout this paper.

The bath-part is also defined by its energy-levels,
\begin{align}
\begin{split}
&H_\mathrm{bath} \ket{E^\mathrm{bath}_j} = E^\mathrm{bath}_j \ket{E^\mathrm{bath}_j} \\
&		E^\mathrm{bath}_{j} = \frac{1}{\beta} \ln \left\{\frac{j}{N} \exp(\beta E^\mathrm{bath}_\mathrm{max})  + (1-\frac{j}{N}) \exp(\beta E^\mathrm{bath}_\mathrm{min})\right\}
\end{split}
\end{align}
while $N$ denotes the dimension of the bath.
This definition yields an (strictly) exponentially growing DOS $\Omega_\mathrm{bath}(E) \propto \exp{\beta E}$ comprising energies from $E^\mathrm{bath}_\mathrm{min}=0$ to $E^\mathrm{bath}_\mathrm{max}=4.5$.
The constant $\beta$ (which takes the role of a temperature here) is chosen to 1.
Note that for this model an exponentially growing DOS of the bath induces an approximately exponentially growing DOS of the full Hamiltonian.

As mentioned in the previous section, an exponentially growing DOS is one of the conditions  (Eq. (\ref{exp_dos})) used here to derive the JR for 
micro-canonical initial states. 
As the  DOS of many physical systems (spatially extended with short range interactions, etc,) is well approximated by an exponential within 
not too large an energy range, this condition is routinely imposed in this context and represents a natural cornerstone of the modelling. Note that 
this modelling corresponds to an "ideal heat bath" i.e., the temperature is always $1/\beta$, regardless of the actual bath energy 
$E^\mathrm{sys}_j$

We now define the interaction between the two parts of system. We introduce the following notation:
\begin{equation}
\ket{E^\mathrm{sys}_m,E^\mathrm{bath}_n} := \ket{E^\mathrm{sys}_m} \otimes \ket{E^\mathrm{bath}_n}
\end{equation}

Regarding this product basis we define the interaction-part.
\begin{align}
\begin{split}
&\bra{E^\mathrm{sys}_m,E^\mathrm{bath}_n} 
   H_\mathrm{int} 
	\ket{E^\mathrm{sys}_k,E^\mathrm{bath}_l} = (1-\delta_{mk})\\
	 & \cdot g(E^\mathrm{bath}_n + E^\mathrm{bath}_l) f( | E^\mathrm{bath}_n - E^\mathrm{bath}_l | ) R_{nl}
\end{split}
\end{align}

\begin{align}
\begin{split}
g(\bar{E}) &= \exp{
	 -\frac{
		 \beta \xi (\bar{E} - E^\mathrm{bath}_\mathrm{max})
	 }
	 {
		 4
	 }}\\
f(\omega) &= 	
   \exp{ 
	- \frac{
		 \omega^2
	 }
	 {
		 2 \sigma_\mathrm{int} ^2
	 }
 }
\end{split}
\end{align}

$R_{nl} = R_{ln}$ denote normally distributed random numbers with zero mean and unit variance.

To assess the rationale behind this modelling consider the following. 

The interaction $ H_\mathrm{int}$ only allows transitions 
(for the non-driven model, i.e., for $\lambda = 0$) between energetically similar bath-states.
Direct transitions between states with significantly different bath-energies are suppressed by the Gaussian function $f(\omega)$, i.e., their suppression is
controlled by the respective  variance $\sigma_\mathrm{int} ^2 = 0.5$.  Within the validity of Fermi's golden rule, 
the decay-rate $\gamma$ of the $z$-component of the 
magnetization of the spin for some  initial bath-energy $ E^\mathrm{bath}$   can be estimated as 
 $\gamma \propto \exp{\beta (1-\xi) E^\mathrm{bath}}$ for our model.
In a physical system we would expect that $\gamma$  depends on the temperature $1/\beta$ of the bath, but not on its actual energy.
For $\xi=1$ the rate $\gamma$    actually becomes independent of the bath energy $E^\mathrm{bath}$. We thus consider this the most physical case.

While it is not plain to be seen, it is an actual and most important fact, that $\xi$ also controls the stiffness of the model. It turns out 
that stiff work-PDF's arise precisely at the above "most physical" case $\xi=1$. For smaller and larger $\xi$ stiffness is lost. For clarity of 
presentation we do not discuss the inner workings of this "stiffness control mechanism" here but simply present clear numerical evidence for its existence
in App. \ref{app_stiffness}.

We finally introduce the time-dependent protocol exclusively acting on the "sys"-part:
\begin{equation}
H_\mathrm{prot}(t) = \sin(\omega_\mathrm{prot} t) (\ket{E^\mathrm{sys}_1}\bra{E^\mathrm{sys}_2} + \mathrm{h.c.})
\end{equation}
Thinking again of the system in terms of a spin-\sfrac{1}{2}-particle, the protocol describes a sinusoidally modulated magnetic field in the $x$-direction,
as routinely used in spin resonance experiments. We choose $\omega_\mathrm{prot} = B_z = 0.5$, i.e., the irradiation is on resonance.
The duration of the protocol is set to $T=3.5 \frac{2 \pi}{\omega_\mathrm{prot}}$ throughout this paper.

\begin{figure}
\includegraphics[width=0.46\textwidth]{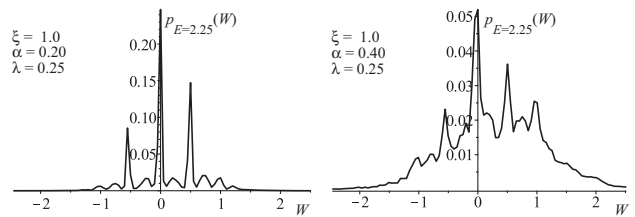}
\caption{
	Work-PDFs for two different bath-couplings $\alpha$.
	For the weaker coupling one nicely sees two sharp peaks at $W=\pm B_\mathrm{z}$,
	    resulting from spin-flips induced by the resonant irradiation.
	For the stronger coupling the work-PDF is much broader.
}
       \label{fig_work_pdfs}
\end{figure}

\section{Jarzynski Relation for Micro-Canonical Initial States and Various System Configurations} \label{numerical_results}

We consider a micro-canonical $\rho^{I_0}_\mathrm{mc}(0)$ initial state from the center of the spectrum of the initial Hamiltonian $H(0)$ with an 
energetic width of about $\delta \approx 0.06$.
\begin{equation} \label{initial_micro}
\rho^{I_0}_\mathrm{mc}(0) = \Omega_{I_0}^{-1} \Pi_{I_0}, \quad I_0 = \Big\lfloor \frac{E_0}{\delta} \Big\rfloor\quad E_0 = \frac{\mathrm{max}(\epsilon_j) + \mathrm{min}(\epsilon_j)}{2}
\end{equation}
The dimension of the bath is set to $N=4000$. 

For this initial state we numerically check the JR for three different stiffness parameters $\xi = 0.6, 1.0, 2.0$ 
with various bath-couplings $\alpha = 0, 0.05, 0.1, \dots, 0.5$ and irradiation-strengths $\lambda = 0, 0.025, 0.05, \dots, 0.25$.

In order to quantify deviations from the perfectly fulfilled JR (Eq. (\ref{jarzynski})) we introduce the following definition:
\begin{eqnarray}
D_\mathrm{mc}(\xi, \alpha, \lambda) &:=& \Tr{ {\bf U}\rho^{I_0}_\mathrm{mc}(0) {\bf U}^+   \exp{-\beta (H(T) - E_0)}} \nonumber \\
&&- \exp{-\beta \Delta F}
\label{jq}
\end{eqnarray}
Since we consider cyclic processes $\Delta F$ is equal to zero and $\exp{-\beta \Delta F}$ becomes equal to 1.
If the JR holds for the considered set of parameters ($\xi$, $\alpha$ and $\lambda$), the corresponding quantifier $D(\xi, \alpha, \lambda)$ vanishes.

\begin{figure}
\includegraphics[width=0.48\textwidth]{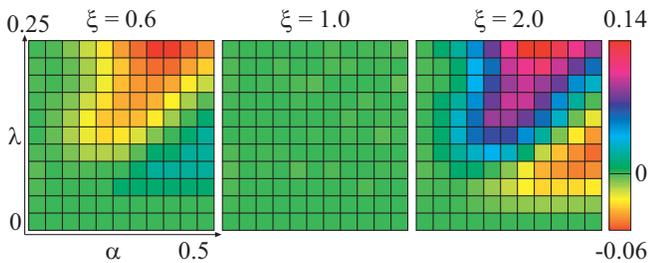}
\caption{$D_\mathrm{mc} (\xi, \alpha, \lambda)$ for various system configurations.
		Light green (zero) indicates that the system complies with the JR, while other colors (non-zero values) quantify the deviations from the JR.
		Apparently the JR is always fulfilled for $\xi = 1$, even for microcanonical initial states.}
       \label{fig_jq}
\end{figure}

The results for the micro-canonical initial states are displayed in Fig. \ref{fig_jq}.
Light green means that the JR is fulfilled, while other colors indicate deviations.

In case of weak bath-couplings $\alpha$ or weak irradiation-strengths $\lambda$ the JR is trivially fulfilled, even for microcanonical initial states.
For $\lambda \approx 0$ we are in the limit of adiabatic following and we thus expect to actually perform zero work.
For $\alpha \approx 0$ the sys- and bath-part are decoupled. But nevertheless the reduced initial sys-state is a thermal
 state with the inverse temperature $\beta$.
So the protocol acts on a system prepared in a Gibbsian state. For this scenario it is well-known that the JR holds.
We therefore concentrate on the larger $\alpha$s and $\lambda$s.

For $\xi=1.0$ the resulting work-PDFs are  stiff, up to small fluctuations (see. App. \ref{app_stiffness}). 
Since stiff work-PDFs imply the JR for micro-canonical initial states, the respective deviations in Fig. \ref{fig_jq} are nearly zero.

For $\xi=0.6, 2.0$ the resulting work-PDFs are not stiff (see. App. \ref{app_stiffness}). 
In principle, the JR could still be fulfilled for microcanonical initial states , since stiffness is formally not a necessary condition. 
However, at both values
i.e, $\xi=0.6, 2.0$, we find deviations from the JR "to both sides" ($D_\mathrm{mc}(\xi, \alpha, \lambda)$ positive as well as negative).
These deviations appear to systematically depend on $\alpha$ and $\lambda$ and are nonzero for most  $\alpha, \lambda$.
However, there are few combinations of $\alpha$ and $\lambda$ for which the JR is fulfilled, see corresponding "light green corridors" in  Fig. \ref{fig_jq}.

In App. \ref{app_energy_dependence} the dependence of the deviations $D_\mathrm{mc}(\xi, \alpha, \lambda)$ on the  initial energy $E_0$ is numerically 
investigated in more detail. 
We find that at $\xi \ne 1$ the initial energy plays a crucial role for the resulting deviations, but not so at $\xi = 1$
Especially at the  "light green corridors" in  Fig. \ref{fig_jq}, left and right panel, the JR is violated for initial microcanonical states with 
energies other than $E_0$.

These numerical finding suggests that the stiffness of work-PDFs is crucial for the validity of the JR
for micro-canonical initial states.

\section{Validity of the Jarzynski relation for Energy Eigenstates and finite size scaling} \label{app_finite_size_scaling}
Up to now we only investigated the validity of the JR for micro-canonical initial states Eq. (\ref{initial_micro}).
We now turn to initial states being eigenstates of the initial Hamiltonian $H(0)$. We denote these initial states as 

\begin{equation}
 \rho^{i}_\mathrm{es}(0) = \ket{i} \bra{i}.
\end{equation}
The energetic width of these states is $\delta=0$. In this sense they are fundamentally different from micro-canonical
initial states. But in this section we will demonstrate, that in the limit of large bath-dimension, both behave similar regarding the JR.

\begin{figure}
\includegraphics[width=0.47 \textwidth]{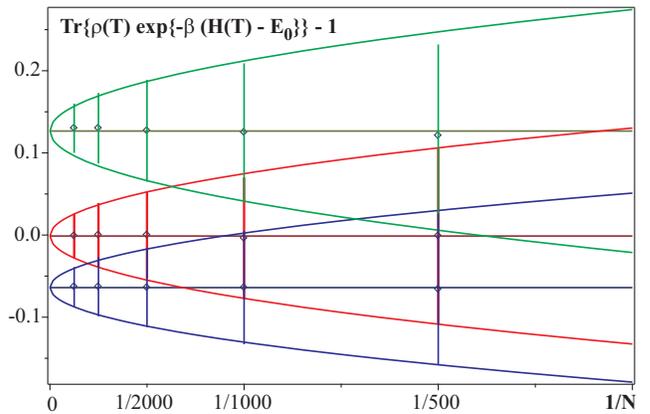}
\caption{
	Finite-size scaling of $D_\mathrm{es}(\xi, \alpha, \lambda)$ for eigenstates (from the center of the spectrum) 
	of the respective initial Hamiltonian $H(0)$.
	Displayed are averages (symbols) and standard-deviations (bars) for three different model parameter sets:
	red: ($\xi = 1$, $\alpha=0.4$, $\lambda=0.25$)
	blue: ($\xi = 0.6$, $\alpha=0.4$, $\lambda=0.25$)
	green: ($\xi = 2.0$, $\alpha=0.4$, $\lambda=0.25$)\\
	{
The standard-deviations are nicely described by tilted parabolae. This suggests that the standard-deviations decrease as $N^{-0.5}$.
	}
}
       \label{finite_size_scaling}
\end{figure}

Again, we use Eq. (\ref{jq}) to check whether the JR is fulfilled or not. We define the corresponding deviations  $D_\mathrm{es}(\xi, \alpha, \lambda)$ 
completely analogous to the  $D_\mathrm{mc}(\xi, \alpha, \lambda)$ (cf. Eq. (\ref{jq})) but with $\rho^{I_0}_\mathrm{mc}(0)$ 
replaced by  $\rho^{i}_\mathrm{es}(0)$
Note that the average of the $D_\mathrm{es}(\xi, \alpha, \lambda)$ over a pertinent range of $i$ equals a corresponding $D_\mathrm{mc}(\xi, \alpha, \lambda)$.
Thus the following numerical results (Fig. \ref{finite_size_scaling}) do not only hold information about the sizes of the 
$D_\mathrm{es}(\xi, \alpha, \lambda)$ but also about the finite size scaling of the $D_\mathrm{mc}(\xi, \alpha, \lambda)$.
   
A systematic survey of the $D_\mathrm{es}(\xi, \alpha, \lambda)$, for all $\alpha, \lambda$ is numerically very costly. We thus concentrate on cases 
where the violation of  the JR is pronounced for $\xi \neq 1$ i.e., $\alpha=0.4$, $\lambda=0.25$, cf. Fig. \ref{fig_jq}.

Figure \ref{finite_size_scaling} shows statistical results on the $D_\mathrm{es}(\xi, \alpha, \lambda)$ for increasing bath sizes $N$. (For clarity 
the results are displayed over inverse bath size $1/N$.) Displayed are the averages (diamonds) and standard deviations (vertical "error" bars) for a
 stiff system $\xi =1$ and two nonstiff systems $\xi=0.6, 2$. The statistics encompass 100 different  $D_\mathrm{es}(\xi, \alpha, \lambda)$ for adjacent $i$ 
 from the middle of the respective spectrum for each parameter set.
   
The following principles may be inferred from Fig. \ref{finite_size_scaling}: The averages appear to be independent of the system size $N$, thus the 
$D_\mathrm{mc}(\xi, \alpha, \lambda)$ are independent of the system size, hence Fig. \ref{fig_jq} provides a representative picture also for other (larger)
bath sizes than $N=4000$. The standard deviations of the  $D_\mathrm{es}(\xi, \alpha, \lambda)$ decrease with bath size, presumably as $\propto N^{-0.5}$ 
as suggested by the tilted parabolae.   

These findings strongly indicate that the JR is indeed fulfilled even for pure initial energy eigenstates for stiff systems in the limit of 
large bath (total system) sizes. Note that in this case the statistical character of the corresponding work-PDFs is entirely due to 
pure quantum uncertainties. Furthermore the JR appears to be always violated for pure   initial energy eigenstates in the limit of 
large bath (total system) sizes if the system is nonstiff.

\section{Discussion}

In this article we analytically show that the Jarzynski relation holds also for a broad class of non-Gibbsian initial states in quantum systems under 
certain conditions. 
For micro-canonical initial states these conditions are :
An exponentially growing DOS of the initial and final Hamiltonian and stiff work-PDF's i.e., work-PDF's that are independent of the initial energy.
Moreover, numerics indicate that the converse also holds: systems that do not comply with the stiffness condition actually do violate the 
JR for micro-canonical initial states, independent of the size of the system.

In order to analytically show the validity of the Jarzynski relation for initial energy eigenstates we 
exploit an additional assumption on the work-PDF's called 
"smoothness", which is expected to hold for large systems.
This expectation is supported by numerics for some examples, which shows that the Jarzynski relation is fulfilled in the limit of large systems for systems 
that do exhibit smoothness, and violated for systems which do not.

To conclude, there appears to be a very tight link between the applicability of the Jarzynski relation and stiffness/smoothness for non-Gibbsian  initial
states which deserves further exploration.

	   {\it Acknowledgements:} 
	   This work has been funded by the Deutsche
	   Forschungsgemeinschaft (DFG) - Grants No. 397107022
	   (GE 1657/3-1), No.
	   355031190 - within the DFG Research Unit FOR 2692.
	   Furthermore this research was supported in part by the National Science Foundation under Grant No. NSF PHY-1748958.

\appendix

\section{Stiffness of Work-PDFs} \label{app_stiffness}
In the main text varied the model parameter $\xi$ and just claimed it would affect the stiffness of the work-PDFs.
In this section we numerically check the actual influence of this model parameter on the work-PDFs.

\begin{figure}
\includegraphics[width=0.47 \textwidth]{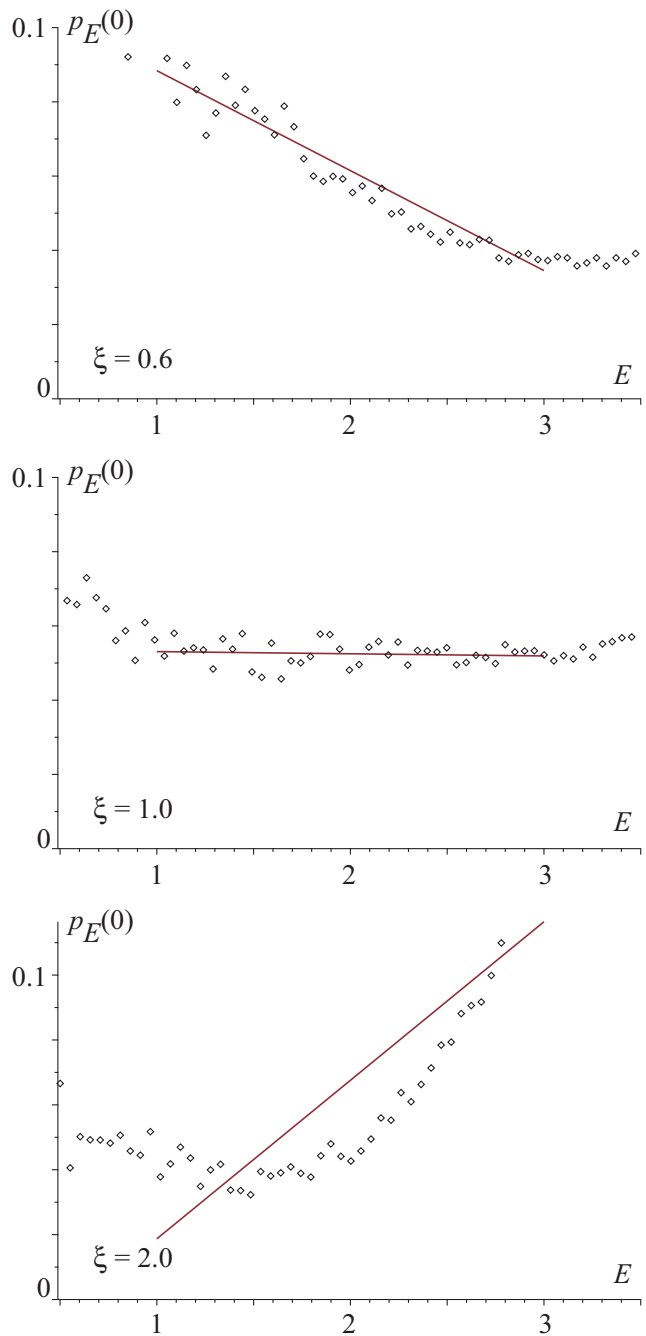}
\caption{
	For $\xi = 1.0$ the probability to perform zero work is approximately independent of the initial energy,
	    while for $\xi = 0.6$ and $\xi = 2.0$ we find a significant dependence.
}
       \label{fig_stiffness}
\end{figure}

We therefore calculated the work-PDFs $p_E(W)$  for various model parameters.
In Fig. \ref{fig_stiffness} we  exemplarily present the data for $d=4000$, $\alpha = 0.4$, $\lambda = 0.25$ and $\xi = 0.6, 1.0, 2.0$ for
eigenstates of $H(0)$ with $E_0 \approx 2.25$. 

Fig. \ref{fig_stiffness} shows the probabilities to perform zero work.
For $\xi=1.0$ the probabilities $p_E(0)$ appear to be approximately independent of $E$, while for $\xi = 0.6$ and $\xi = 2.0$
we find a significant dependence.

While for larger bath dimensions $d$ the work-PDFs become smoother, the slope for $\xi=0.6, 2.0$ appears to be independent of $d$.

\section{Jarzynski Realtion for different initial energies} \label{app_energy_dependence}
\begin{figure}
\includegraphics[width=0.47 \textwidth]{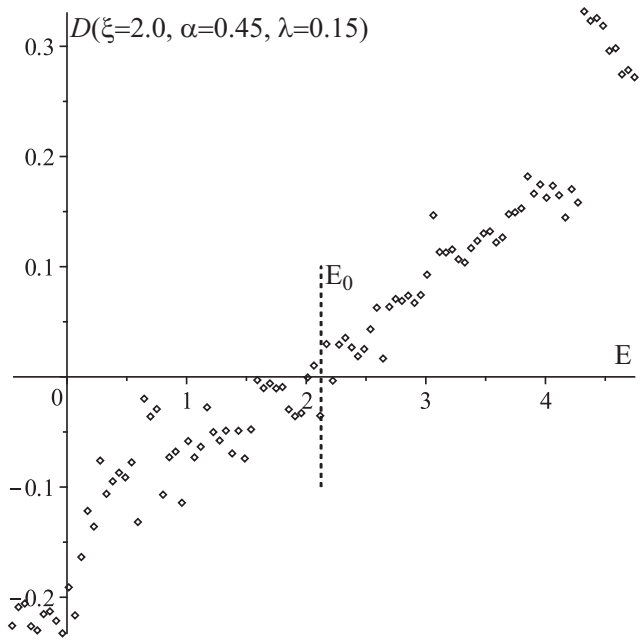}
\caption{
	Energy dependence of $D(\xi, \alpha, \lambda)$.
}
       \label{fig_energy_dependence}
\end{figure}

In Sec. \ref{numerical_results} we considered deviations from the JR for various combinations of $\xi$, $\alpha$ and $\lambda$, but for a fixed initial energy $E_0$ and found
that for some combinations of these parameters the JR appeared to be fulfilled, even though condition Eq. (\ref{stiffness}) is violated.
We now consider the dependence of these deviations on the energy of the initial state $\rho(0)$ with the aforementioned parameters held constant.
We consider micro-canonical initial states, defined according to Eq. (\ref{initial_micro}), with various energies $E$.
The resulting deviations $D(\xi=2.0, \alpha=0.45, \lambda=0.15)$ are displayed in Fig (\ref{fig_energy_dependence}).

Note that for this parameter combination we found the JR fulfilled for the previously considered initial energy $E_0$.
The data suggests that there is only a small energy-range for which the JR is approximately fulfilled and $E_0$ accidentally is within this region.
The energy-dependence for other $\alpha$ and $\lambda$ looks quite similar. So we can find specific micro-canonical initial states,
	which comply with the JR, even if condition Eq. (\ref{stiffness}) is not fulfilled. 
But since this is a feature of a very specific combination of system and initial state we conclude that the JR is not fulfilled by this system and driving-protocol in general.

In contrast, for $\xi=1$ there is a wide region of initial energies that fulfill the JR, which is a direct consequence of the conditions Eq. (\ref{stiffness}) and Eq. (\ref{exp_dos}).

\bibliography{mybib,refs}{}

\end{document}